\documentstyle[12pt]{article}
\def\singlespace {\smallskipamount=3.75pt plus1pt minus1pt
                  \medskipamount=7.5pt plus2pt minus2pt
                  \bigskipamount=15pt plus4pt minus4pt
                  \normalbaselineskip=15pt plus0pt minus0pt
                  \normallineskip=1pt
                  \normallineskiplimit=0pt
                  \jot=3.75pt
                  {\def\smallskip {\vskip\smallskipamount}}
                  {\def\medskip   {\vskip\medskipamount}}
                  {\def\bigskip   {\vskip\bigskipamount}}
                  {\setbox\strutbox=\hbox{\vrule
                    height10.5pt depth4.5pt width 0pt}}
                  \parskip 7.5pt
                  \normalbaselines}
\def\middlespace {\smallskipamount=5.825pt plus1.5pt minus1.5pt
                  \medskipamount=11.25pt plus3pt minus3pt
                  \bigskipamount=22.5pt plus6pt minus6pt
                  \normalbaselineskip=22.5pt plus0pt minus0pt
                  \normallineskip=1pt
                  \normallineskiplimit=0pt
                  \jot=5.825pt
                  {\def\smallskip {\vskip\smallskipamount}}
                  {\def\medskip   {\vskip\medskipamount}}
                  {\def\bigskip   {\vskip\bigskipamount}}
                  {\setbox\strutbox=\hbox{\vrule
                    height15.75pt depth6.75pt width 0pt}}
                  \parskip 7.25pt
                  \normalbaselines}
\def\dblspc {\smallskipamount=7.5pt plus2pt minus2pt
                  \medskipamount=15pt plus4pt minus4pt
                  \bigskipamount=30pt plus8pt minus8pt
                  \normalbaselineskip=30pt plus0pt minus0pt
                  \normallineskip=2pt
                  \normallineskiplimit=0pt
                  \jot=7.5pt
                  {\def\smallskip {\vskip\smallskipamount}}
                  {\def\medskip   {\vskip\medskipamount}}
                  {\def\bigskip   {\vskip\bigskipamount}}
                  {\setbox\strutbox=\hbox{\vrule
                    height21.0pt depth9.0pt width 0pt}}
                  \parskip 15.0pt
                  \normalbaselines}

\def\be{\begin{equation}}

\def\j-{\J_-}

\def\ee{\end{equation}}

\def\bearr{\begin{eqnarray}}
\def\bearrs{\begin{eqnarray*}}
\def\eearr{\end{eqnarray}}
\def\eearrs{\end{eqnarray*}}
\def\barr{\begin{array}}
\def\earr{\end{array}}

\def\non\non{\nonumber}
\def\nn8{\nonumber\\[15pt]}
\def\l{\left}
\def\r{\right}
\def\un{\underline}
\def\ve{\varepsilon}
\def\f{\frac}

\def\dis{\displaystyle}

\begin{document}
\thispagestyle{empty}
\begin{center}
{\Large { Astrophysical signals of P and T
violation by gravity
}}\\[20pt]

{\bf A.R. Prasanna and Subhendra  Mohanty \\
Physical Research Laboratory\\
Ahmedabad 380 009, India}\\[30pt]

ABSTRACT\\
\end{center}

\dblspc{
We study the observational consequences of the leading order P
and T violating gauge invariant graviton-photon interactions.
Such interactions give rise to gravitational birefringence - the
velocity of light signals depends upon their polarisation, which
is absent in Einstein's gravity.
Using  experimentally established limits on the
differential time delay in the radio signals from pulsars we put
 constraints on the magnitude of P and T violating
graviton-photon couplings.\\

\newpage
It is well eastablished experimentally that P(parity) and CP
(charge conjugation and parity) symetries are violated by the
weak interactions \cite{one}. The CP violation in the
electroweak theory can be accounted for by the complex phases
of the Kobayashi-Masakawa  quark matrix. In theories like QED
where there is no tree level CP violation it is expected that
radiative corrections due to the quark loops will give rise to
CP violating dimension five operators like a fermion electric
dipole momment \cite{two}. Experimental limits on electron and neutron
electric dipole momments provide strong constraints on the CP violating
extensions of the standard model \cite{three}.
A good motivation to test for  CP violating gravitational
interactions is that these may arise from radiative corrections
to graviton-matter vertices due to  some CP
odd couplings in the matter sector \cite{four}. Putting
experimental bounds on CP odd (which is equivalent to P and T
odd , if CPT is conserved) graviton-matter couplings can lead
to bounds on the magnitude of CP violation present in the
 extensions of the standard model.
A second reason to look for
the presence of CP violating operators in gravitational
interactions is that in the
full quantum theory of gravity
such operators may be present at the tree level and there may be measurably
large violation of CP in gravitational interactions on which
there is so far no strong experimental constraints.

In this letter we consider the leading order P and T violating graviton-photon
interactions  which are
 invariant under the $U(1)_{EM}$ and the general coordinate
transformations . The P and T odd  operator of dimension four
$F^{\mu\nu}{\tilde F}^{\mu \nu}$ is a total divergence and
therefore plays no dynamical role.
The leading order CP violating graviton-photon interaction
Lagrangian is of dimension six and is of the form
\be
\barr{lll}
L& =& \sqrt{-g} \l\{ F_{\mu \nu} F^{\mu \nu} + c_1 R_{\mu \nu
\alpha \beta} F^{\mu \nu} \tilde{F}^{\alpha \beta}\right.\\[8pt]
&+&\left. c_2 R_{\mu \nu} F^{\mu \alpha} {\tilde{F}^\nu}_\alpha + c_3 R
F_{\mu\nu} \tilde{F}^{\mu\nu}\r\}
\earr
\ee
where $\tilde{F}^{\mu\nu} \equiv (1/2)
\ve^{\alpha\beta\mu\nu} ~F_{\alpha \beta}$ and the unknown
 coefficients $c_i$ have dimensions of $M^{-2}$.
We show that the addition of such higher derivative terms to Einsteins
gravity leads to the phenomenon of birefringence i.e photons of different
polarisations follow different trajectories in a background Kerr
metric. We show that the presence of such an interaction would
give rise to a differential time delay between the positive and
negative circular polarisations of pulsar signals. Observations
of signals from the pulsar PSR 1937+21 \cite{cordes} place an upper
bound of $10^{-6} sec$ on the differential time delay between
the circularly polarised modes \cite{loseco}. This enables us to
place an upper bound on one of the CP violating couplings of
(1), namely $c_1 < 10^{-9} sec^2$.

 The
equation of motion obtained from the interactions (1)
is given by
\be
\barr{lll}
\nabla_\mu F^{\mu\nu}&+& c_1 \l[ \l( \nabla_\mu
{R^{\mu\nu}}_{\alpha \beta} \r) \tilde{F}^{\alpha \beta} +
{R^{\mu\nu}}_{\alpha \beta} \l( \nabla_\mu \tilde{F}^{\alpha
\beta} \r) \r]\\[8pt]
&+& c_2 \f{1}{2} \l[ \nabla_\mu \l( {R^\nu}_\alpha \tilde{F}^{\alpha
\mu} - {R^\mu}_\alpha \tilde{F}^{\alpha \nu} \r) \r]\\[8pt]
&+&c_3 \l[ R \nabla_\mu \tilde{F}^{\mu \nu} + \tilde{F}^{\mu
\nu} \nabla_\mu R \r] = 0
\earr
\ee
Using the Bianchi identity
\be
\nabla_\mu {R^{\mu \nu}}_{\alpha\beta} = \nabla_\alpha {R^\nu}_\beta
- \nabla_\beta {R^\nu}_\alpha
\ee
we can write the equations of motion  (2) in the form
\be
\barr{lll}
\nabla_\mu F^{\mu \nu} &+& c_1 {R^{\mu\nu}}_{\alpha\beta}
\nabla_\mu \tilde{F}^{\alpha\beta} + \l( 2c_1 + c_2 \f{1}{2} \r)
\l( \nabla_\alpha {R^\nu}_\beta \r) \tilde{F}^{\alpha \beta}\\[8pt]
&+& c_2 \f{1}{2} \l[ \l( \nabla_\mu {R^\mu}_\alpha \r)
\tilde{F}^{\alpha \nu} + {R^\mu}_\alpha \l( \nabla_\mu
\tilde{F}^{\alpha \nu} \r) \r]\\[8pt]
&+&c_3 \l( \nabla_\mu R \r) \tilde{F}^{\mu \nu} = 0
\earr
\ee
For propagation of photons in the Kerr metric, where
$R_{\mu\nu} = R = 0$, the equations of motion (4) reduce to
the form
\be
\nabla_\mu F^{\mu\nu} + c_1 {R^{\mu\nu}}_{\alpha\beta}
{}~~\ve^{\alpha\beta\gamma\delta} \nabla_\mu F_{\gamma\delta} = 0
\ee
In terms of the gauge potential $A_\mu$ one can write (5) as the
wave equation
\be
\nabla_\mu \nabla^\mu A^\nu + 2 c_1 {R^{\mu\nu}}_{\alpha\beta}
{}~~\ve^{\alpha\beta\gamma\delta} \nabla_\mu \nabla_\delta A_\gamma
= 0
\ee
where we have chosen the gauge $\nabla^\mu A_\mu = 0$. The
photon trajectories can be obtained from the wave equation (6)
by making the eikonal ansatz
\be
A^\nu = exp \l(\f{iS}{\ve} \r) {\dis \sum}_n \l( \f{\ve}{i}
\r)^n \hat{A}^\nu \equiv expo \l( \f{iS}{\ve} \r) f^\nu
\ee
Substituting (7) in (6) and collecting the various powers of
$(1/\ve )$ we obtain the following set of  equations
$$
\f{-1}{\epsilon^2} \l\{ \l( \partial_\mu S \r) \l( \partial^\mu S \r)
\hat{A}^\nu + 2c_1 {R^{\mu\nu}}_{\alpha\beta}
{}~~\ve^{\alpha\beta\gamma\delta} \l( \partial_\delta S \r) \l(
\partial_\mu S \r) \hat{A}_\gamma \r\} = 0\eqno(8a)
$$
$$\f{i}{\ve} \l\{ 2 \l(\partial_\mu S \r) \nabla^\mu f^\nu +
\nabla^\mu \l( \partial_\mu S \r) f^\nu \right. \\[8pt]
+ 2c_1 {R^{\mu\nu}}_{\alpha\beta} ~~\ve^{\alpha\beta\gamma\delta}
\l\{ \l[ \l( \partial_\mu S \r) \nabla_\delta + \l( \partial_\delta
S \r) \nabla_\mu \r] f_\gamma\right.\\
\left. + \l( \nabla_\mu \partial_\delta S \r) f_\gamma \r\} = 0\eqno(8b)
$$
$$\nabla_\mu \nabla^\mu f^\nu + 2c_1 {R^{\mu\nu}}_{\alpha\beta}
{}~~\ve^{\alpha\beta\gamma\delta} \l( \nabla_\mu \nabla_\delta
f_\gamma \r) = 0\eqno(8c)
$$
\addtocounter{equation}{1}
Identifying the wave vector $k_\mu \equiv \partial_\mu S$, we see
that (8a) is the dispersion relation which gives the trajectory
of the ray, (8b) gives the equation for the parallel transport
of the polarisation vector $f_\nu$ and gives the Faraday
rotation in the curved background, and (8c) gives the intensity
along the light trajectory.

  We assume that space-time exterior to a pulsar
is described by the  linearised Kerr metric with components
\be
\barr{lll}
g_{00}&=& - \l( 1 - \f{2GM}{r} \r) = - g^{-1}_{rr}\\
g_{0\phi}&=& - \f{2GJ}{r} \sin^2\theta\\
g_{\theta\theta}&=& r^2 = g_{\phi\phi} \l( \sin^2\theta \r)^{-1}
\earr
\ee
where $J$ is the angular momentum per unit mass.  Using (I9)
to compute the components of the Riemann tensor
${R^{\mu\nu}}_{\alpha\beta}$ and substituting in (8a) we get the
disperson relations for the spatial components of the four
potential $A_\nu$ (we choose the gauge $A_0 = 0$) given by
\be
\l( \barr{ccc}k^2&\beta\omega^2&0\\
\beta\omega^2&k^2&0\\0&0&k^2\earr \r) \l(
\barr{c}A_r\\A_\theta\\A_\phi \earr \r) = 0
\ee
where
\be
\beta= 12 c_1 \f{GJ}{r^4} (- g_{00})^{-3/2}
\ee
and
\be
k^2=   g^{00} \omega^2 + g_{0\phi}
k_\phi  \omega + g^{rr} k^2_r
+g^{\theta\theta} k^2_\theta + g^{\phi\phi} k^2_\phi
\ee
For radial trajectories with $k_\theta = k_\phi = 0$ we can use
the gauge condition
\be
g^{rr} k_r A_r + g^{0\phi} \omega A_\phi = 0
\ee
to obtain from (10) equation for the transverse components $\l(
A_\theta, A_\phi \r)$,
\be
\l( \barr{cc}\beta \omega^2&\alpha \gamma\\ \alpha&\beta \gamma \omega^2
\earr \r) \l( \barr{c}A_\theta\\ A_\phi \earr \r) = 0
\ee
where
$$
\alpha = g^{rr} k^2_r + g^{00} \omega^2\eqno(14a)$$
and $$\gamma = \f{g^{0\phi} \omega}{g^{rr} k_r}\eqno(14b)$$
\addtocounter{equation}{1}
The dispersion relation is obtained by setting the determinant
of the matrix in (13) equal to zero to give
\be
g^{rr} k^2_r + g^{00} \omega^2 \mp \beta \omega^2 = 0
\ee
which can be solved for the eigenvalues
\be
k_r^\pm =  \omega (\f{-g^{00}}{g^{11}}) \l( 1 \pm
\f{\beta}{2g^{00}} \r)
\ee
for the propagating modes which in this case are the
elliptically polarised modes $\l( 1 + \ve^2 \r)^{-1/2} \l(
A_\theta \pm \ve A_\phi \r)$.
The propagation time of the $\pm$ modes is given by the
Hamilton-Jacobi condition
\be
t^\pm = \int \f{\partial k^\pm}{\partial \omega} dr
\ee
Substituting for $k^\pm_r$ from (16) we find that the
differential time delay between the two modes is given by
\be
\delta t = t^+ - t^- = \int^D_R \f{\beta}{(-g_{00}g^{11})^{1/2}} dr
=  c_1~ 4 GJ
\l[ \f{1}{R^3} - \f{9GM}{4R^4} \r]
\ee
A typical millisecond pulsar has angular momentum $J \simeq 10^{48} gm~
cm^2 ~sec^{-1}$,  mass $M \simeq 1.4 M_{\odot}$ and
 radius $R \simeq 10~ km$.  From the
experimental constraint \cite{cordes,loseco}
that the observed differential time
delay between the $+$ and $-$ polarised modes $\delta t <
10^{-6} sec$, we can, using (18), put an upper bound on the
coefficient $c_1$
\be
c_1 < 10^{-9} sec^2
\ee
The other couplings in (1) with coefficients $c_2$ and $c_3$
would be
relevant when the wave propagation is through dense matter  such that
$R^{\mu \nu}$ are $R$ are nonzero. However in such situations it would
be hard to separate the birefringence due to matter from the birefringence
due to gravity.

 Einstein's theory predicts that, massive test particles with
non-zero spin or angular momentum, in an external gravitation
field, follow geodesics which  depend upon the orientation of
the spin-angular momentum \cite{papa}. It has been claimed \cite{mash}
 that such
an effect also holds for photons
in Einsteins gravity
if one solves the wave equation beyond the eikonal approximation.
However, we have shown \cite{pras1} that according to the exact solution
of the wave equation in curved space, photons of all
polarisations follow null geodesics and thus there cannot be any
birefringence in Einstein gravity.  Gravitational
birefringence appears only due to non-minimal coupling as was
shown by Bedram and Lesche \cite{bedram} , using the non-minimal
couplings introduced by Prasanna \cite{pras} , which can arise
from the radiative corrections to Einstein's gravity \cite{drum,shore}.
 CP violatations
occuring in  particle physics models can give rise to CP violating
graviton-photon couplings \cite{four} through loop corrections.
In this paper we have discussed the experimental signatures of such
couplings and put constriants on their magnitude from
observational data.

\newpage



\begin{thebibliography}{99}
\bibitem{one} H. Christenson et al Phys Rev Lett 13, 138 , (1964).

\bibitem{two} E. P. Shabalin, Sov J Nucl Phys 28, 75 (1978) ;
F. Hoogeven, Nucl Phys. B 341, 322 (1990).

\bibitem{three} Abdullah et al , Phys Rev Lett 65, 2347, (1990).

\bibitem{four} S. Mohanty and S. D. Rindani, PRL preprint, 1995.


\bibitem{cordes} Cordes, J.M. and Stinebring, D.R., Ap.L. Lett.
\un{277}, 53 (1984).

\bibitem{loseco} Loseco, J.M., et.al., Phy. Lett. \un{A138}, 5 (1989);
Klien, J.R. and Thorsett, S.E., Phy. Lett. \un{A145}, 79 (1990).

\bibitem{papa} Papapetrou A., Proc. R. Soc. London \un{A209},
248-258 (1951).

\bibitem{mash} Mashoon, B., Nature \un{250}, 316-317 (1974);
{\it ibid}, Phys. Rev. \un{D7}, 2807-2814 (1973).

\bibitem{pras1} Mohanty S. and  Prasanna A.R, Preprint Astro-ph 9505100.

\bibitem{bedram} Bedram, M.L. and Lesche, B., (1986) GR11
Abstract Book, No. 03:08, p. 137, Stockholm Conference on GRG.

\bibitem{pras} Prasanna A.R. Phys Lett 37A, 331, (1971); Lett
Nuovo Cim. 6 (11), 420 (1973).

\bibitem{drum} I.T. Drummond and S.J Hathrell,
Phys. Rev D22 (1980)343.

\bibitem{shore} R.D. Daniels and G.M.Shore,
Nucl. Phys. B425, (1994) 634;G.M. Shore, Preprint gr-qc 9504041.

\end{thebibliography}
\end{document}